\documentclass{PoS}

\title{Improving Constraints on Proton Structure using CMS measurements}

\ShortTitle{Improving Constraints on Proton Structure using CMS measurements}

\author{\speaker{Saranya Ghosh}\thanks{On behalf of the CMS Collaboration}\\
        TIFR\\
        E-mail: \email{saranya.ghosh@cern.ch}}

\abstract{Production of electroweak bosons, heavy quarks and jets in proton-proton collisions probe different aspects of QCD and are sensitive to the details of proton structure, expressed by parton distribution functions (PDFs). Precise measurements of cross sections of these processes are used by the CMS experiment to demonstrate the impact of the LHC data on the PDFs and their precision. The measurements of muon charge asymmetry in W-boson production at a center-of-mass of 7 and 8 TeV is used to improve the constraints on the valence-quark distributions, while the associated production of W-boson and charm quark provides information on the s-quark distribution in the proton. Production of inclusive jets, as measured by CMS at center-of-mass energy of 7 TeV, provides important constraints on the gluon distribution.}

\FullConference{The European Physical Society Conference on High Energy Physics\\
		22--29 July 2015\\
		Vienna, Austria}

\begin{document}

\section{Introduction}
\label{sec1Intro}

Parton Distribution Functions (PDFs) contain information about the structure of hadrons. Hadrons contain quarks and gluons, together referred to as partons. The PDFs give the probability of a particular parton carrying a certain fraction of the of the total momentum of the hadron. A good understanding of the PDFs is essential for theoretical predictions of the cross section of any physics process at hadron colliders, such as the Large Hadron Collider (LHC), where the cross section is given by :

\begin{equation}
\sigma_{X}(Q^{2}) = \sum_{ij}{\int \int dx_{i}dx_{j}f_{i}(x_{i}, Q^{2})f_{j}(x_{j}, Q^{2})\hat{\sigma}_{ij \rightarrow X}({x_{i},x_{j},Q^{2})}}        
\end{equation}

Here, $Q^{2}$ is the square of the momentum scale of the process, the summation is for all parton pairs within the colliding hadrons, integrations are over the full kinematic phase-space and $x_{i} = { p_{i}}/{ p}$ is the parton momentum fraction with ${ p}$ and ${ p_{i}}$ being the momentum of the hadron and the momentum carried by the $i^{th}$ parton inside the proton respectively. The $\hat{\sigma}_{ij\rightarrow X}$  is the cross section at the parton level. $f_{i}(x_{i},Q^{2})$ refers to the PDF of the $i^{th}$ parton, and it is a function of the momentum fraction ($x$) carried by the parton and momentum scale ($Q^{2}$) of the physics process.  

\begin{figure}[htbp]
\centering
\includegraphics[scale=.25]{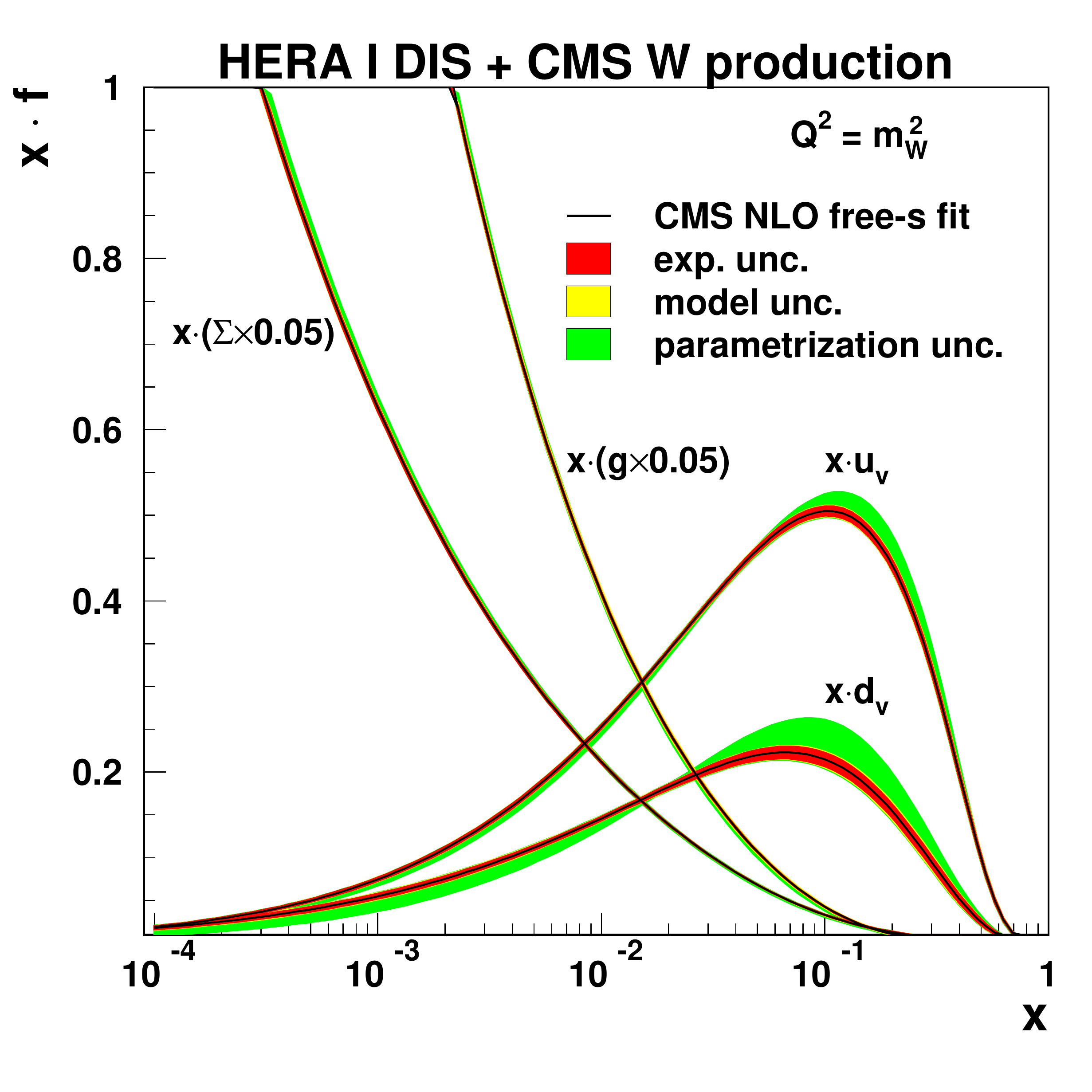}
\caption{
Proton PDFs based on  HERA DIS data and CMS measurements of W-boson production shown as a function of $x$ at $Q^{2} = m_{W}^{2}$ \cite{bib:CMSMuonaAsym7}.
}
\label{fig:protonpdf}
\end{figure}

PDFs are usually evaluated by parametrising the shape dependence on the momentum fraction ($x$) at low values of the momentum scale ($Q^{2}$) using fits to
data from mainly Deep Inelastic Scattering (DIS) experiments. The PDFs are then evolved to higher $Q^{2}$ using theory while the $x$ dependance has to be obtained from fits to experimental measurements. Precise measurements performed by the Compact Muon Solenoid (CMS) experiment \cite{bib:CMS} at the LHC are sensitive to PDFs and these measurements are used to improve the precision of the of the PDFs. Proton PDFs based on  HERA DIS data and CMS measurements of W-boson production are shown in Figure~\ref{fig:protonpdf}. 

\section{CMS measurements that are sensitive to PDFs}
\label{sec2CMSmeasure}

Different CMS measurements that are sensitive to PDFs are discussed below.

\subsection{W differential lepton charge asymmetry}
\label{subsec:wasym}

\begin{figure}[htbp]
\centering
\includegraphics[scale=.3]{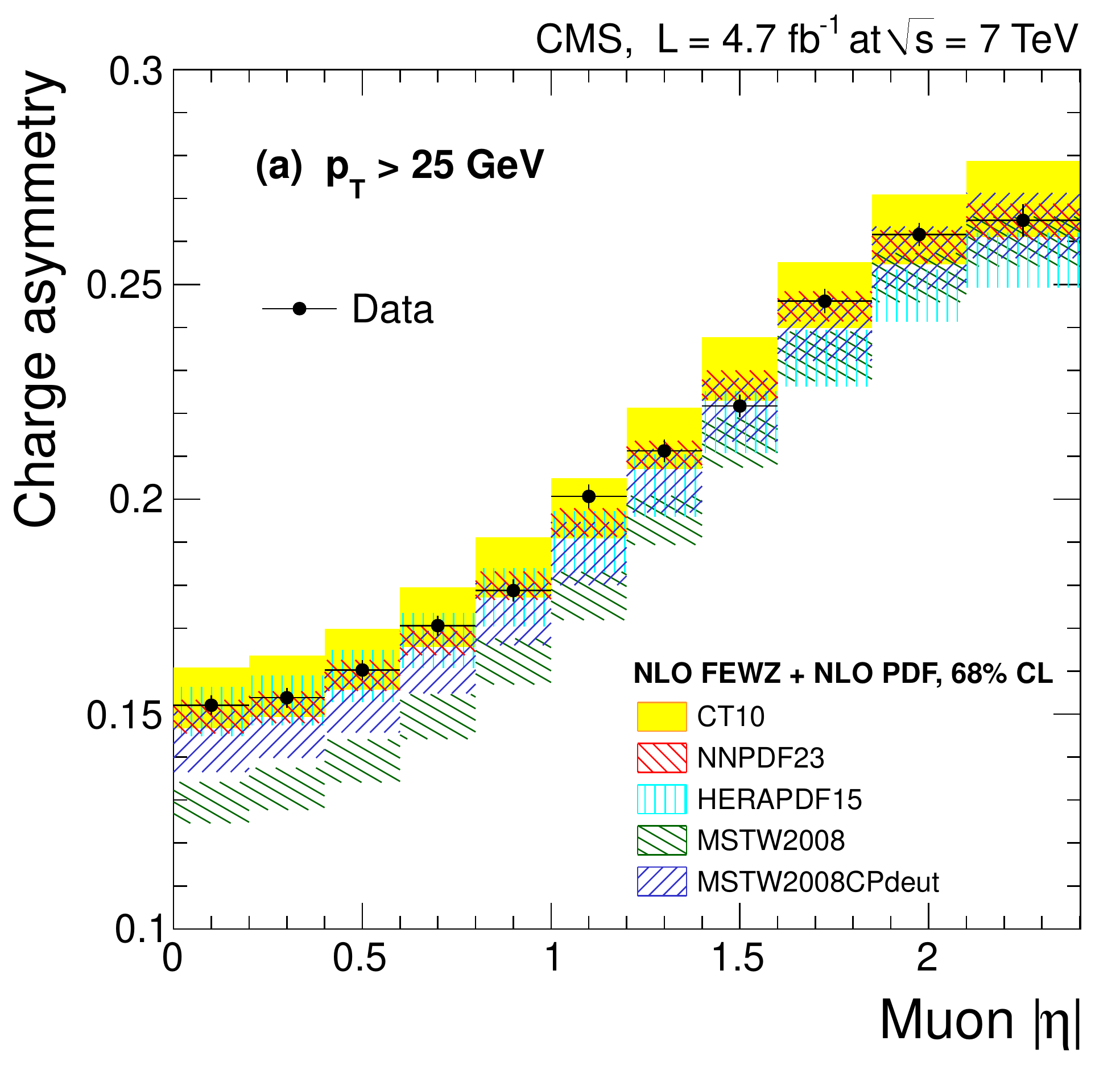}
\caption{
Comparison of the measured muon charge asymmetry at $\sqrt{s}$ = 7 TeV to the NLO predictions calculated using the FEWZ 3.1 MC tool interfaced with the NLO CT10, NNPDF2.3, HERAPDF1.5, MSTW2008, and MSTW2008CPdeut PDF sets \cite{bib:CMSMuonaAsym7}.
}
\label{fig:muonasym}
\end{figure}

According to the standard model, the leading order processes for inclusive $W$-boson production in the proton-proton collisions at the LHC are the annihilation processes: $u\bar{d} \rightarrow W^{+}$ and $d\bar{u} \rightarrow W^{-}$, where a valence quark comes from one proton and a sea quark comes from the other proton.
Measurements of the asymmetry in the production of $W^{+}$ and $W^{-}$ bosons as a function of the boson rapidity are sensitive to the PDFs of the valence $u$ and $d$ quarks and also the sea quark densities. It is difficult to measure the asymmetry in terms of the boson rapidity because of the energy taken away by the neutrinos in leptonic $W$-boson decays and hence the measurement of the decay lepton charge asymmetry is performed as a function of the pseudorapidity of the lepton. The differential lepton charge asymmetry from the leptonic decay of $W$-bosons is defined as :

\begin{equation}
\mathcal{A}(\eta) = \frac{\frac{d\sigma}{d\eta}(W^{+}\rightarrow\ell^{+}\nu) - \frac{d\sigma}{d\eta}(W^{-}\rightarrow\ell^{-}\bar{\nu}) }{\frac{d\sigma}{d\eta}(W^{+}\rightarrow\ell^{+}\nu) + \frac{d\sigma}{d\eta}(W^{-}\rightarrow\ell^{-}\bar{\nu})}        
\end{equation}

where $d\sigma/d\eta$ is the differential cross section for $W$-boson production and the subsequent leptonic decay and $\eta$ = -ln[tan($\theta$/2)] is the pseudorapidity of the decay lepton in the laboratory frame, with $\theta$ being the polar angle measured with respect to the beam axis.
\\
\\
Measurements of the lepton charge asymmetry have been performed by the CMS experiment in the electron channel using 840 $pb^{-1}$ of data collected with proton-proton collisions at $\sqrt{s}$ = 7 TeV \cite{bib:CMSElecAsym} and in the muon channel using 4.7 $fb^{-1}$ of data collected at $\sqrt{s}$ = 7 TeV \cite{bib:CMSMuonaAsym7} and 18.8 $fb^{-1}$ of data collected at $\sqrt{s}$ = 8 TeV \cite{bib:CMSMuonaAsym8}. The measurement of the muon charge asymmetry as $\sqrt{s}$ = 7 TeV is shown in Figure~\ref{fig:muonasym} along with comparisons to the NLO predictions of the asymmetry calculated using four different PDF sets. 

\subsection{W + c differential cross section}
\label{subsec:wc}

\begin{figure}[htbp]
\centering
\includegraphics[scale=.55]{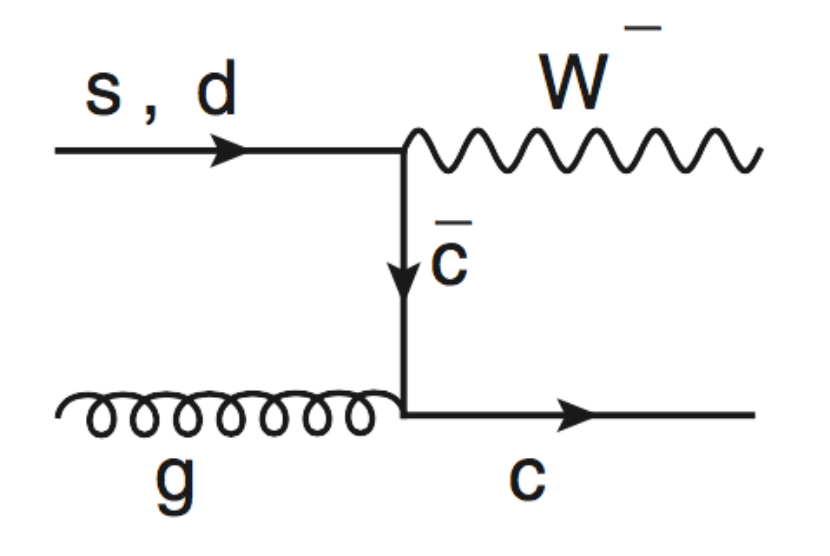}
\includegraphics[scale=.55]{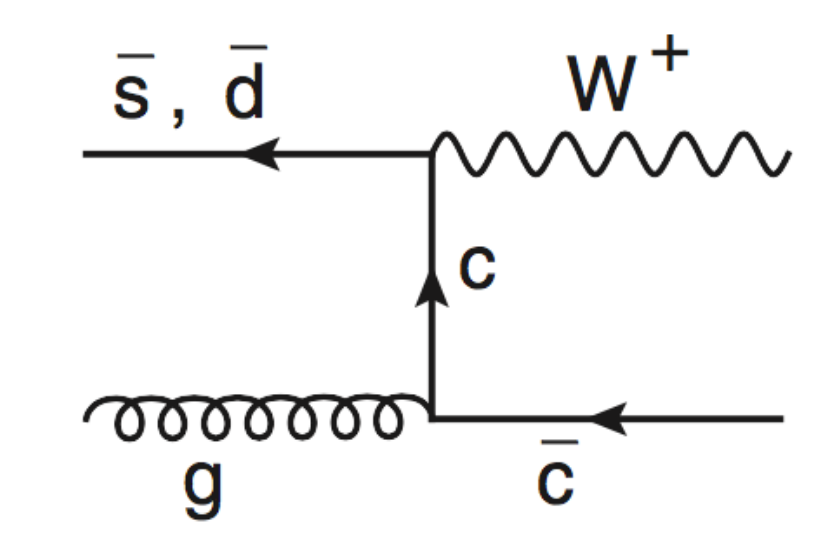}
\caption{
Main diagrams for the associated W + c production at the LHC.
}
\label{fig:W+cFeyn}
\end{figure}

Studies regarding the production of a W boson in association with a charm ($c$) quark jet, together referred to as W + c production, at the LHC can provide access to the strange quark and anti-quark content of the proton. This is due to the dominance of the $sg\rightarrow W^{-}+c$ and $\bar{s}g\rightarrow W^{+}+\bar{c}$ contributions to W + c production, with the contributions from the down ($d$) quark being Cabibbo suppressed. The main Feynman diagrams  at the hard-scattering level for associated W + c production at the LHC are shown in Figure~\ref{fig:W+cFeyn}. 

\begin{figure}[htbp]
\centering
\includegraphics[scale=.3]{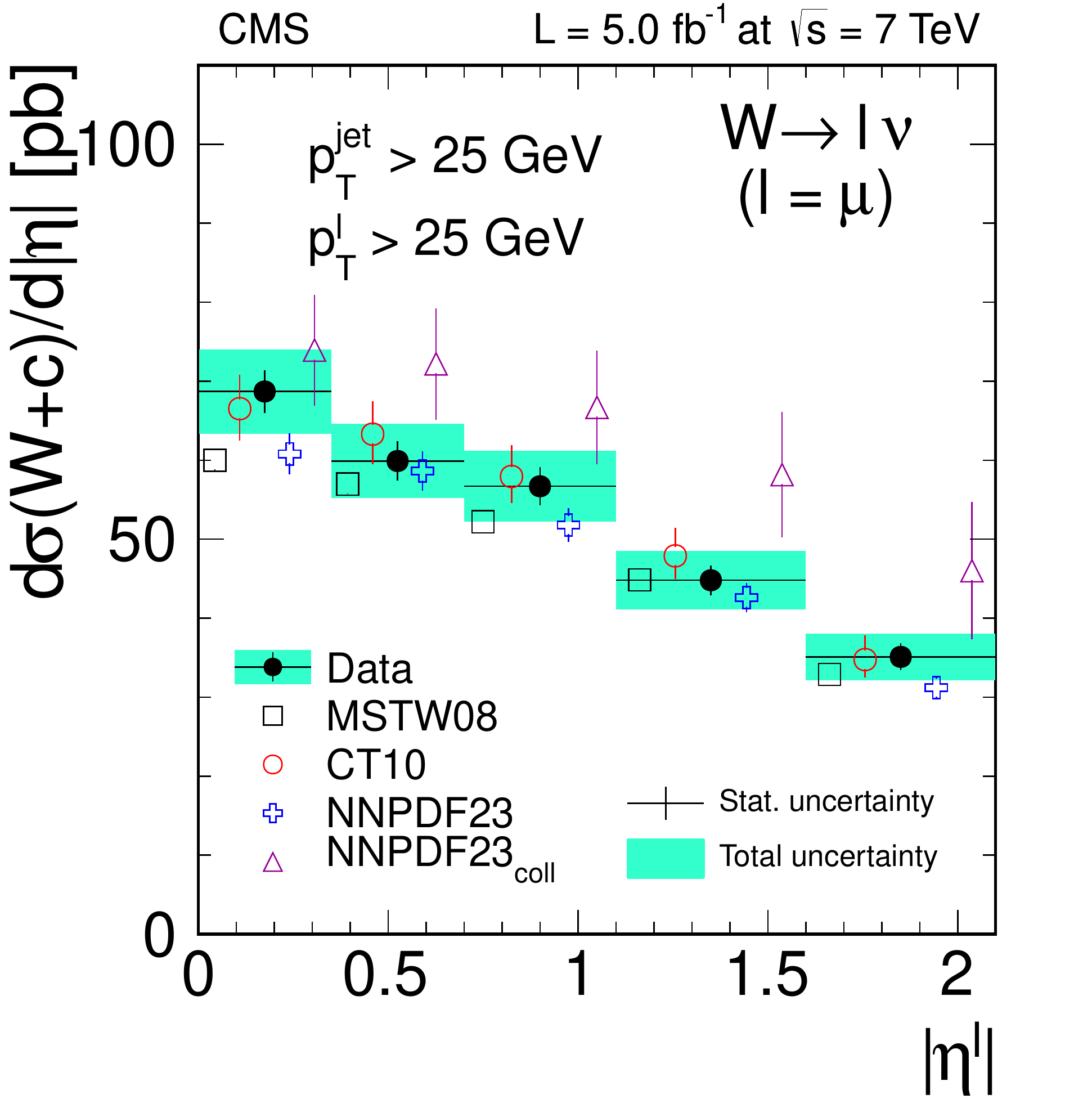}
\caption{
Differential cross section, $d\sigma(W+c)/d|\eta|$, as a function of the absolute value of the pseudorapidity of the lepton from the W-boson decay, compared with the theoretical predictions at NLO computed with MCFM and four different PDF sets, MSTW2008, CT10, NNPDF2.3 and NNPDF2.3$_{coll}$  \cite{bib:CMSW+c}.
}
\label{fig:W+cresult}
\end{figure}

Measurements of the differential cross sections of the associated W + c production with respect to the absolute value of the pseudorapidity of the lepton from the $W$-boson decay ($d\sigma(W+c)/d|\eta|$) have been performed by the CMS experiment in the electron and the muon channel of $W$-boson decay using a data sample corresponding to a total integrated luminosity of 5 $fb^{-1}$ collected by the CMS detector at the LHC during proton-proton collisions at $\sqrt{s}$ = 7 TeV \cite{bib:CMSW+c}. A comparison of the measurement for the muon channel with predictions from different PDF sets is shown in Figure~\ref{fig:W+cresult}.

\subsection{Inclusive jet cross section}
\label{subsec:incjetcsn}

\begin{figure}[htbp]
\centering
\includegraphics[scale=.26]{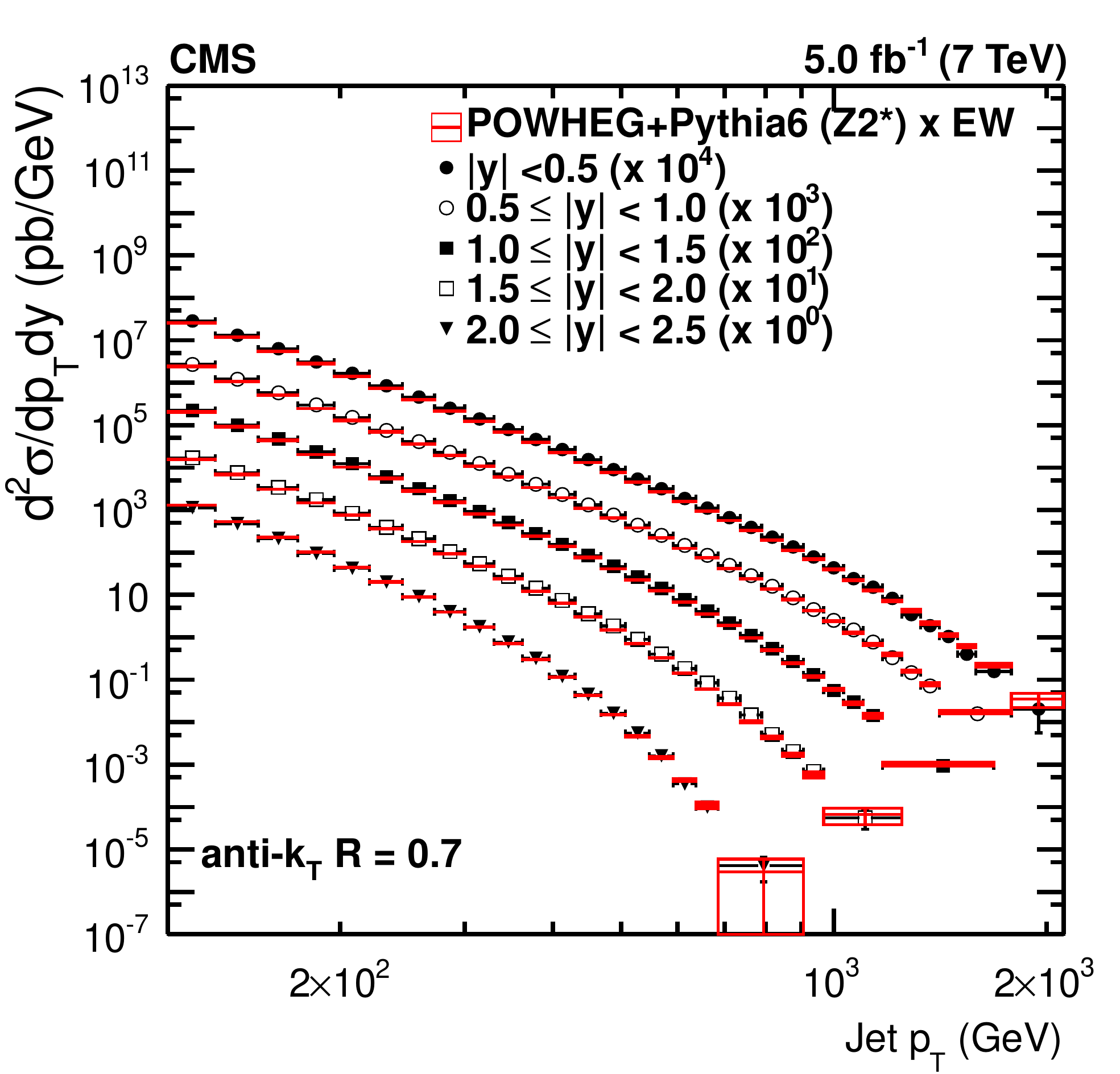}
\includegraphics[scale=.29]{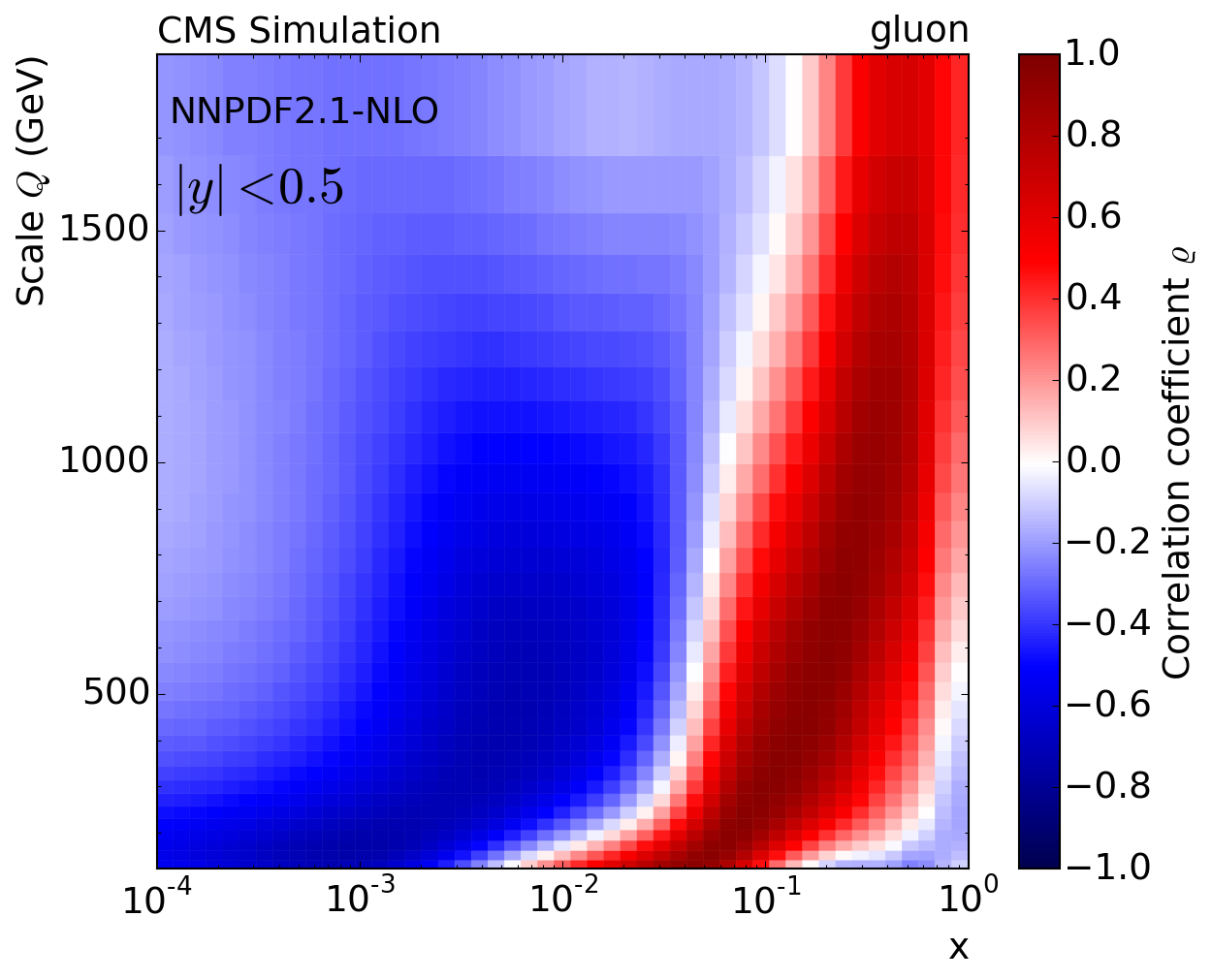}
\caption{
The double-differential inclusive jet cross section in comparison to NLO predictions using the NNPDF2.1 PDF set \cite{bib:incJetMeasurement} (left) and the correlation coefficient between the inclusive jet cross section and the gluon PDF as a function of the momentum fraction $x$ of the proton and the momentum scale $Q$ of the process for the central rapidity region \cite{bib:incJetPDFAnalysis} (right).
}
\label{fig:jetGluonCorln}
\end{figure}

Measurement of the inclusive jet cross section in proton-proton collisions, double differential in jet transverse momentum ($p_{T}$) and absolute rapidity ($|y|$) has been performed by the CMS experiment \cite{bib:incJetMeasurement}. 
The CMS measurement of the inclusive jet cross section is shown on the left in Figure~\ref{fig:jetGluonCorln}. The inclusive jet cross section measurement can be used to constrain PDFs, especially that for the gluons. Studies performed by CMS to evaluate the correlation between the inclusive jet cross section and the PDFs of the different partons \cite{bib:incJetPDFAnalysis} show strong correlation with the gluon PDF at high $x$ as shown in Figure~\ref{fig:jetGluonCorln} on the right. 

\section{Impact of CMS measurements on PDFs}
\label{sec3CMS_PDFs}

The HERAFITTER framework \cite{bib:HERAFITTER} is an open source framework that is designed to fit PDFs to data, along with other applications. The HERAFITTER framework was used to fit the proton PDFs to data using CMS measurements along with DIS data from HERA \cite{bib:HERA} to demonstrate the impact of the CMS measurements on constraining the PDFs. The measurement of the W differential lepton charge asymmetry has a significant impact on the valence quark PDFs, while the associated W + c differential cross section measurements impacts the PDFs of the strange quark and antiquark and the inclusive jet cross section measurements provides strong constraints on the gluon PDFs. 

\begin{figure}[htbp]
\centering
\includegraphics[scale=.18]{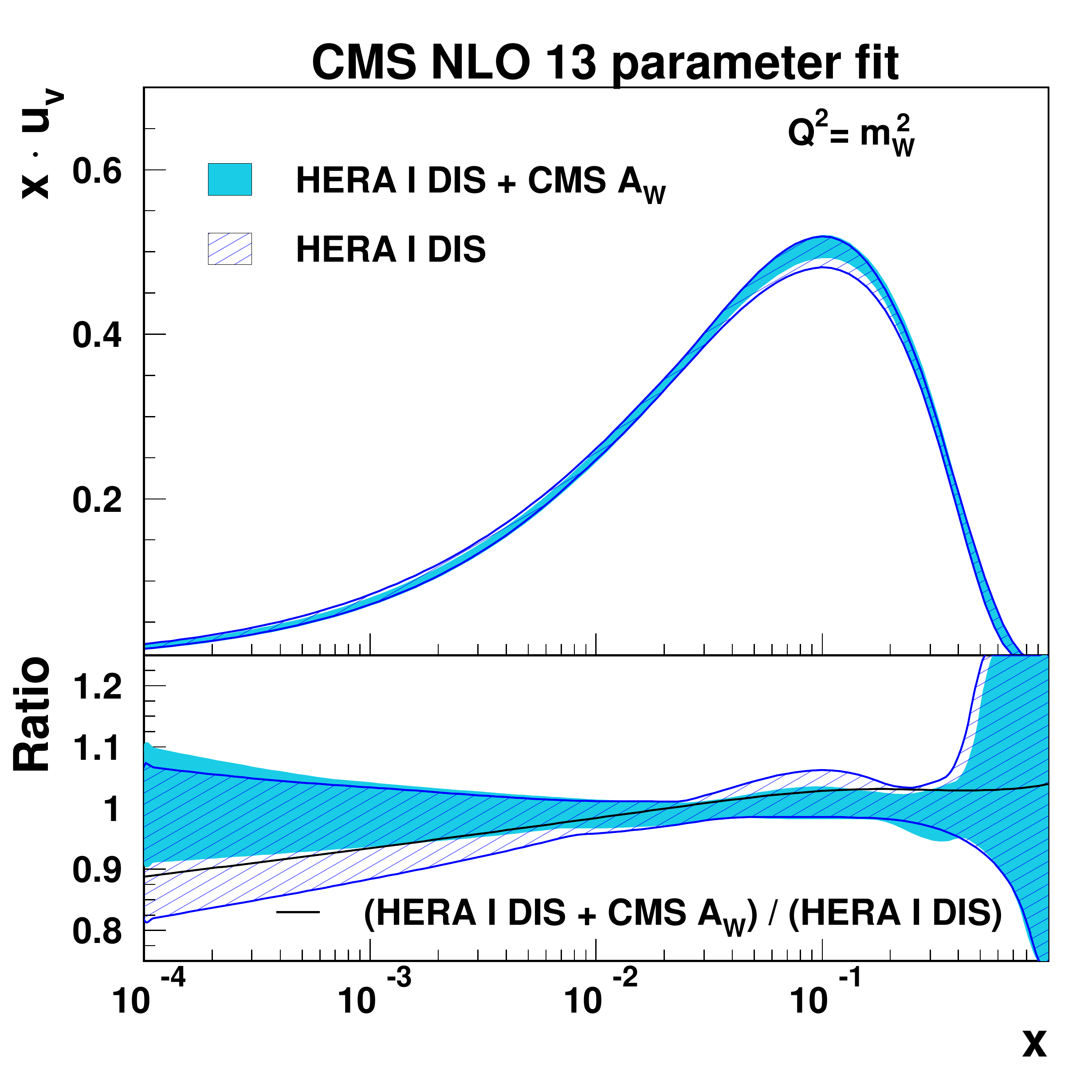}
\includegraphics[scale=.18]{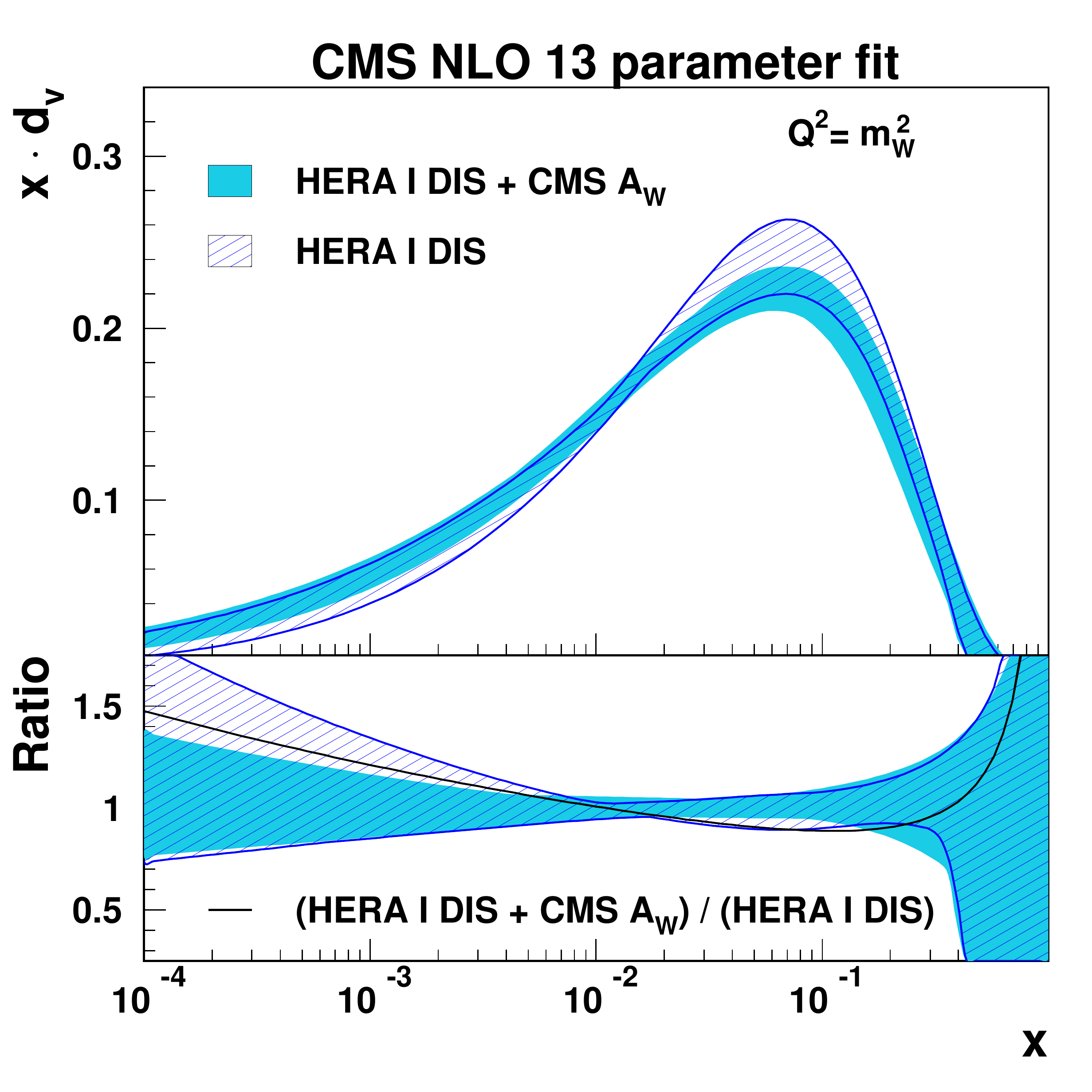}
\includegraphics[scale=.28]{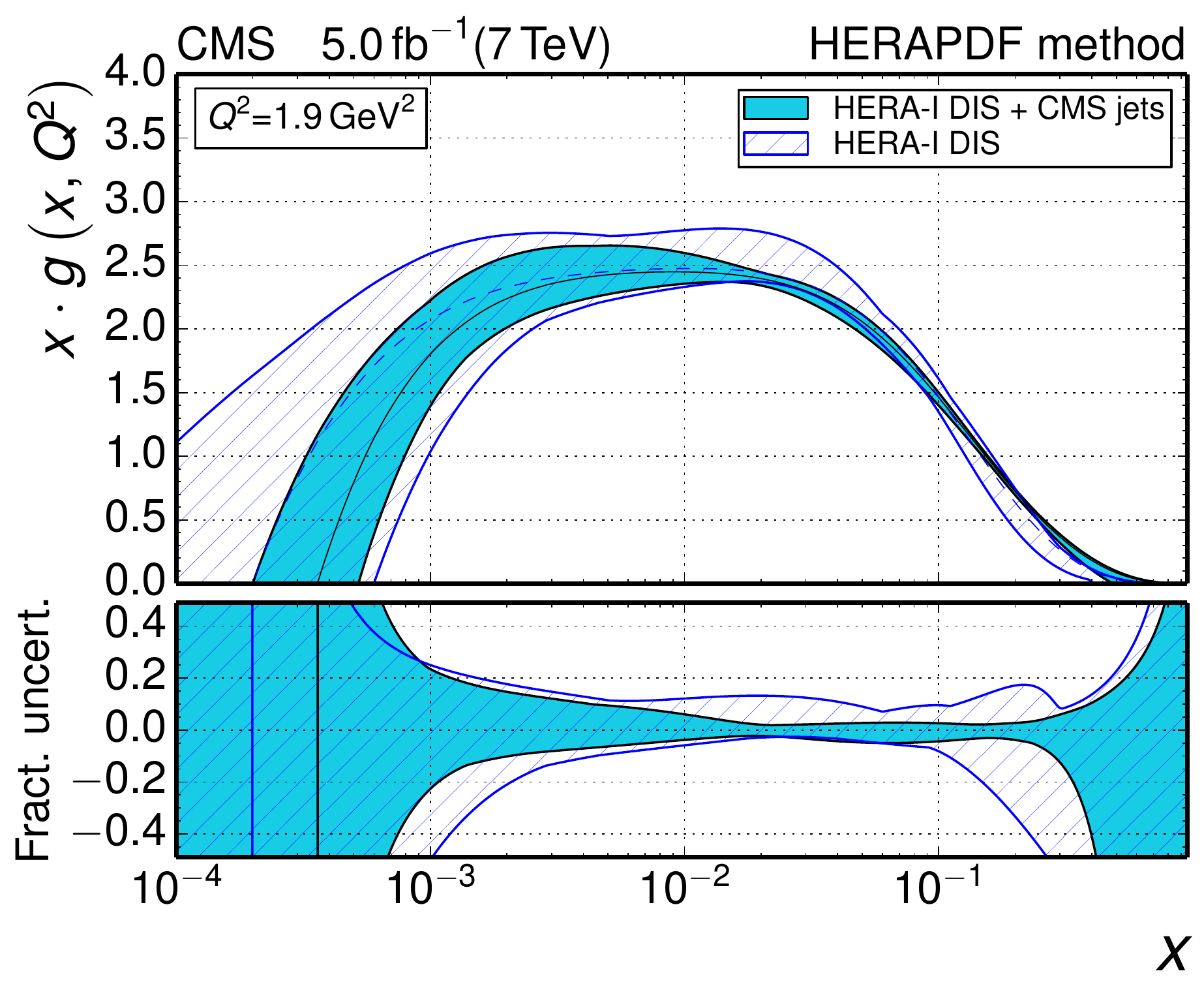}
\caption{
Distributions of u valence (left), d valence (center) quarks \cite{bib:CMSMuonaAsym7} and gluons \cite{bib:incJetPDFAnalysis} (right) as functions of $x$ at the scale $Q^{2} = m^{2}_{W}$ (1.9 $GeV^{2}$ for gluons). The results of the fit to the HERA data and CMS measurements (light shaded band), and to HERA only (dark hatched band) are compared. In the bottom panels the change of the PDFs with respect to the fits using HERA data only is represented by the solid line, along with the normalised distributions for a direct comparison of the uncertainties.
}
\label{fig:pdfuv}
\end{figure}

\begin{figure}[htbp]
\centering
\includegraphics[scale=.22]{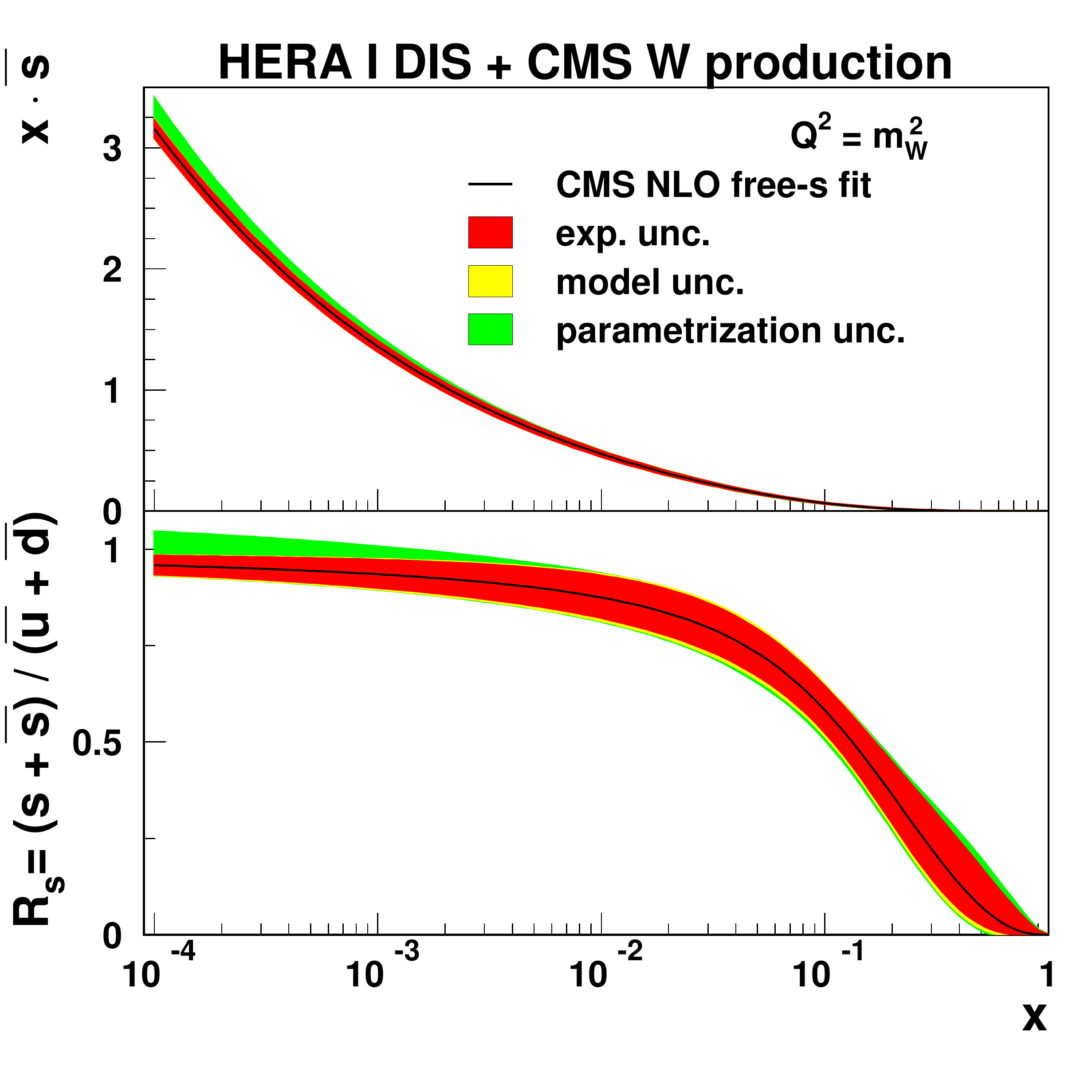}
\caption{
Strange antiquark distribution ($\bar{s}(x,Q)$) and the ratio of the distributions of the strange quarks to that of the up and down antiquarks ($R_{s}(x,Q)$), obtained using HERA and CMS data, shown as functions of $x$ at the scale $Q^{2} = m^{2}_{W}$  \cite{bib:CMSMuonaAsym7}.
}
\label{fig:pdfstrange}
\end{figure}

Figure~\ref{fig:pdfuv} shows the valence up ($u_{v}$) and down ($d_{v}$) quark PDFs evaluated using HERA data only compared to the PDFs evaluated using CMS measurement of the W differential muon charge asymmetry at $\sqrt{s}$ = 7 TeV in combination with the HERA data along with a similar comparison of the gluon PDF using HERA data alone with that using the CMS inclusive jet cross section measurement at $\sqrt{s}$ = 7 TeV along with the HERA data. These figures show that the CMS measurements have a significant impact on the PDFs. Figure~\ref{fig:pdfstrange} shows the PDF for the strange antiquark evaluated using the CMS W + c measurement and HERA data, and also the ratio of the PDFs of the strange quarks to the up and down antiquarks that are part of the sea quarks in the proton.

\section{Conclusion}
\label{sec4Conc}

Measurements of different standard model processes performed by the CMS experiment have been shown to have significant potential in terms of constraining the PDFs of valence quarks, gluons and sea quarks in the proton. These measurements can be used in global PDF fits to improve our understanding of the proton structure.

\end{document}